\documentclass[times, twoside]{zHenriquesLab-StyleBioRxiv}
\usepackage{blindtext}
\usepackage{lipsum}
\leadauthor{Kotte} 

\begin{document}
\title{MIDOG 2025 Track 2: A Deep Learning Model for Classification of Atypical and Normal Mitotic Figures under Class and Hardness Imbalances}
\shorttitle{Modelling class and hardness imbalances}

\author[a]{Sujatha Kotte}
\author[a]{Vangala Govindakrishnan Saipradeep}
\author[a]{Vidushi Walia}
\author[a]{Dhandapani Nandagopal}
\author[a]{Thomas Joseph}
\author[a]{Naveen Sivadasan}
\author[b]{Bhagat Singh Lali}

\affil[a]{TCS Research, Tata Consultancy Services (TCS) Ltd,, Hyderabad, India}
\affil[b]{Department of Pathology, Tata Medical Center, Kolkata, India}

\maketitle

\begin{abstract}
Motivation: Accurate classification of mitotic figures into normal and atypical types is crucial for tumor prognostication in digital pathology. However, developing robust deep learning models for this task is challenging due to the subtle morphological differences, as well as significant class and hardness imbalances in real-world histopathology datasets.\hfill\break
Methods: We propose a novel deep learning approach based on a ResNet backbone with specialized classification heads. Our architecture uniquely models both the mitotic figure phenotype and the instance difficulty simultaneously. This method is specifically designed to handle the challenges of diverse tissue types, scanner variability, and imbalanced data. We employed focal loss to effectively mitigate the pronounced class imbalance, and a comprehensive data augmentation pipeline was implemented to enhance the model's robustness and generalizability.\hfill\break
Results: Our approach demonstrated strong and consistent performance. In a 5-fold cross-validation on the MIDOG 2025 Track 2 dataset, it achieved a mean balanced accuracy of 0.8744±0.0093 and an ROC AUC of 0.9505±0.029. The model showed robust generalization across preliminary leaderboard evaluations, achieving an overall balanced accuracy of 0.8736±0.0204.\hfill\break
Conclusion: The proposed method offers a reliable and generalizable solution for the classification of atypical and normal mitotic figures. By addressing the inherent challenges of real-world data, our approach has the potential to support precise prognostic assessments in clinical practice and improve consistency in pathological diagnosis.

\end {abstract}


\begin{corrauthor}
saipradeep.v@tcs.com
\end{corrauthor}

\section*{Introduction}
Mitosis is a fundamental phase of the cell cycle during which a parent cell divides to form two daughter cells which are genetically identical, facilitating tissue growth and cellular turnover. In cancer, dysregulation of the cell cycle leads to aberrant mitotic activity, contributing to uncontrolled proliferation \cite{mitosis}. Observation of unusual, dysregulated, and random assembly of the contents of the nucleus in dividing cells results in abnormal mitotic morphology - referred to as atypical mitosis. This could be due to some genomic abnormalities like telomere dysfunction, aneuploidy or chromosomal instability, among others. The aggressiveness of cancers is often positively correlated with high count of atypical mitotic figures. Presence of atypical mitotic figures is suggested to be of prognostic value in some tumors like pancreatic and urothelial carcinomas \cite{AtypMitBC}.

The standard practice in histopathology for identifying mitotic figures involves manual inspection of hematoxylin and eosin (H\&E) stained histological sections. With advancements in digital pathology and high-resolution whole-slide scanners \cite{DigPathMitosis}, there has been significant progress in digitizing histological samples for computational analysis. However, many current deep learning models are trained on WSIs from a single tumor type, resulting in decreased performance when applied across diverse cancer types. Additionally, variability in staining protocols and scanner hardware across laboratories introduces further inconsistencies, which negatively affect the precision and recall of automated detection and classification systems. These limitations have hindered widespread clinical adoption of computational pathology tools \cite{Limitations}.

There have been initiatives to help address the problems related to automated detection of mitotic figures in the form of challenges such as TUPAC16 \cite{TUPAC16}, Mitos \& Atypia 14 \cite{mitos}, MIDOG-2021 \cite{midog2021} and MIDOG-2022 \cite{midog2022}. Building on these, the MIDOG-2025 Track-2 challenge \cite{MIDOG2025} aims to help develop models for the accurate classification of atypical mitotic figures and normal mitotic figures across a wide range of tissue regions. The models should be able to handle the variability in tissue morphology across histological samples, in addition to the morphological overlap between mitotic figures and other structures. The complexity is even more pronounced due to the subtle morphological differences between atypical mitotic figures from normal mitotic figures, which is a difficult to discern even for experienced pathologists. Other factors include the tissue types, tumor vs non-tumor regions, scanners etc. The methods developed should enable accurate classification, which is critical in understanding biology of tumors and as a consequence, improve prognostic assessments when faced with the variability and complexity of real-world histopathological samples.

To this end, we propose our approach for robust and accurate classification of typical and atypical mitotic figures.

\section*{Material and Methods}

\subsection{Data}\hfill\break
Subtype information distinguishing normal from atypical mitotic figures is rarely available in publicly accessible datasets. To overcome this limitation, the MIDOG 2025 challenge introduced a subtype dataset derived from MIDOG++ \cite{midog++}. This dataset consists of 128×128 pixel cropped cell patches, each labeled by expert pathologists as either normal or atypical mitotic figures. In total, 10,191 patches are annotated as normal and 1,748 as atypical, sourced from 454 labeled whole-slide images (WSIs). These WSIs encompass 10 domains, each representing a unique combination of tumor types, species (human or canine), scanners, and laboratories, providing a diverse range of imaging conditions. 

The training set consists of samples from seven tumor types, spanning two species and acquired using five different scanners across four laboratories. The distribution of data is uneven, with 41.9\% of patches originating from VMU, Vienna, and only 9.1\% from AMC, New York. Human tissue samples constitute 29.4\% of the dataset, with the remainder from canine samples; no feline samples are included. The 3D Histech scanner contributed slightly over 51\% of the training dataset. Notably, class imbalance is present, with atypical mitotic figures (AMFs) comprising 14.65\% (n = 1,748) and normal mitotic figures (NMFs) 85.35\% (n = 10,191) of the total dataset. 

Annotations were performed independently by three expert pathologists, and a majority consensus label was assigned to each patch. Disagreement among annotators was observed in 1,639 patches (13.7\%), comprising 575 NMFs and 1,064 AMFs; these are designated as “hard” instances (suffix ‘H’), while patches with full agreement are denoted as “easy” (suffix ‘E’). 

\subsection{Model Architecture}\hfill\break
To address the inherent diversity, class imbalance, and presence of challenging instances within the MIDOG 2025 subtype dataset, we implemented a deep convolutional neural network based on the ResNet architecture. A comprehensive series of ablation studies were conducted, comparing various convolutional neural networks (CNNs), transformer-based models, pathology foundation models, and hybrid architectures. The ResNet50 backbone emerged as the optimal choice, demonstrating both computational efficiency in terms of model size and computational speed with competitive performance as measured by balanced accuracy and F1 score. 

The proposed model architecture consists of the following components: (a) a ResNet-based CNN backbone initialized with ImageNet-pretrained weights; (b) binary classification heads for each expert’s annotation, distinguishing atypical mitotic figures (AMFs; negative label, 0) from normal mitotic figures (NMFs; positive label, 1); (c) a binary classification head to differentiate “hard” (0) versus “easy” (1) instances, based on consensus among annotators; and (d) a shared fully connected layer interfacing the backbone with the aforementioned classification heads, thereby modeling both phenotype and instance difficulty simultaneously. ResNet50 was selected in preference to deeper variants including ResNet101 and ResNet152, given its superior efficiency and equivalent classification metrics. 

To mitigate the pronounced class imbalance with respect to both AMF/NMF labels and “hard”/“easy” instances, focal loss was employed as the primary loss function. The focal losses from the expert classification and hardness heads are combined with (theta=0.5) to calculate final loss.

The ResNet-50 backbone extracts a 2048-dimensional feature vector from input 128X128 image patches which is fed to a shared fully connected layer of 512 dimensions and subsequently projected to binary classification heads for AMF, NMF classification and hard, easy instance classification.

\subsection{Cross Validation}\hfill\break
A stratified 5-fold cross-validation protocol was implemented on the training dataset to maintain consistent distributions of atypical mitotic figures (AMFs) versus normal mitotic figures (NMFs), as well as “easy” versus “hard” instances across all folds. This approach facilitates robust model assessment and generalization, addressing data imbalance while minimizing the risk of overfitting. 

\subsection{Augmentations}\hfill\break
To address staining heterogeneity introduced by various scanners and morphological diversity present within normal mitotic figures (NMFs) and atypical mitotic figures (AMFs), we implemented a comprehensive augmentation pipeline. The augmentations consisted of: 
\begin{enumerate}
    \item Resizing images to 224×224 pixels, 
    \item Random horizontal flipping, 
    \item Random vertical flipping, 
    \item Random rotations within ±10 degrees, 
    \item Adjustment of image brightness (±0.2), contrast (±0.3), and saturation (±0.1) using ColorJitter, 
    \item Normalization according to ImageNet statistics. 
\end{enumerate}

Collectively, these transformations were aimed to enhance the model's robustness to inter-domain variability and morphological discrepancies, thereby supporting generalizable feature extraction across the dataset. 

\subsection{Setup}\hfill\break
Training was performed using the AdamW optimizer with a cosine annealing learning rate scheduler, starting from an initial learning rate of $1\times10^{-4}$ with a batch size of 8. The model checkpoints were selected from the best-performing epochs, identified through early stopping metric based on maximum validation balanced accuracy with a patience of 20. 

Hyperparameters for focal loss (alpha and gamma), learning rate, dropout were systematically optimized using Optuna  hyperparameter optimization framework.

\section*{Experiments and Results}

Table 1 summarizes the performance metrics of the proposed method obtained through 5-fold stratified cross-validation and preliminary evaluation on the leaderboard, which included four tumor types. Our approach demonstrated consistent results across cross-validation folds, achieving a mean balanced accuracy of 0.8744 (±0.0093). Comparable performance was observed in the preliminary leaderboard evaluation, with an overall balanced accuracy of 0.8736 (±0.0204) across the four tumor categories, indicating robust generalization to unseen tumour types in the preliminary evaluation data. 

In addition, the method attained an overall area under the receiver operating characteristic curve (ROC AUC) of 0.9505 (±0.029), underscoring its capacity to discriminate between atypical mitotic figures (AMFs) and normal mitotic figures (NMFs). Sensitivity analysis revealed that, for three of the four tumor types, the model achieved a perfect sensitivity score of 1.0, corresponding to an absence of false negative predictions in those categories. 

We evaluated a modified classification approach in which the binary hardness labels (“easy” and “hard”) were expanded into four distinct categories: NMF Easy, NMF Hard, AMF Easy, and AMF Hard. This finer-grained labeling resulted in a decrease in balanced accuracy of 0.0148 as compared with the original two-class hardness scheme. 

\textbf{Context and Stain Models}: Additional experiments were conducted to enhance model performance by integrating features from both stain-based and image-based domains. Specifically, Hematoxylin, Eosin, and DAB (HED) channels were extracted and combined with standard RGB channels to enrich the input feature set. To further capture local context, a secondary patch (80 × 80 px) was cropped from within the original 128 × 128 px region, providing supplementary contextual information surrounding the mitotic figures. Results from five-fold cross-validation indicated that the combination of both contextual/cropped RGB and HED patches outperformed models utilizing only original or cropped regions, yielding a mean balanced accuracy of 0.8854 (±0.0069). 

\begin{table}[ht!]
\centering
\setlength{\tabcolsep}{3pt}
\begin{tabular}{|l|c|c|c|c|}
\hline
\multicolumn{1}{|c|}{\textbf{Dataset}} & \textbf{\begin{tabular}[c]{@{}c@{}}Balanced \\ Accuracy\end{tabular}} & \textbf{Specificity}                                       & \textbf{Sensitivity}                                       & \textbf{ROC AUC} \\ \hline
\textbf{Validation}                    & \begin{tabular}[c]{@{}c@{}}0.8744\\ (±0.0093)\end{tabular}            & \begin{tabular}[c]{@{}c@{}}0.9032\\ (±0.0157)\end{tabular} & \begin{tabular}[c]{@{}c@{}}0.8676\\ (±0.0166)\end{tabular} & -                \\ \hline\hline 
\textbf{Domain 1}                      & 0.8906                                                                & 0.7813                                                     & 1.0                                                        & 0.9375           \\ \hline
\textbf{Domain 2}                      & 0.8463                                                                & 0.7273                                                     & 0.9655                                                     & 0.9089           \\ \hline
\textbf{Domain 3}                      & 0.8989                                                                & 0.7978                                                     & 1.0                                                        & 0.9696           \\ \hline
\textbf{Domain 4}                      & 0.8889                                                                & 0.7777                                                     & 1.0                                                        & 0.9861           \\ \hline
\end{tabular}
\caption{Performance results of our proposed approach on the Validation dataset (first row) and Preliminary evaluation test of MIDOG 2025 - Track 2 }
\label{tab:my-table}
\end{table}

We have performed extensive ablation studies for choosing the right architecture for our approach using 5-fold cross validation.

\textbf{Backbones} We systematically evaluated multiple model architectures, including conventional convolutional neural networks (CNNs) such as EfficientNet and ResNet (including deeper variants EfficientNet-b7 and ResNet152), as well as transformer-based models such as SwinTransformer. In addition, linear probing was performed using the pathology foundation model Provg-Gigapath. Comparative analyses demonstrated that traditional CNNs consistently outperformed both the linear-probed Provg-Gigapath foundation model and transformer-based architectures. For instance, linear probed Provg-Gigapath was able to achieve a balanced accuracy of 0.70 when compared to ResNet50 (0.8585), ResNet101 (0.8634) and EfficientB0 (0.8449) respectively. 

\textbf{Hybrid models} Hybrid modeling strategies, which incorporated frozen foundation model features in conjunction with pretrained convolutional neural network (CNN) backbones, were evaluated for their classification performance. Notably, these hybrid configurations did not surpass the results obtained from fully fine-tuned CNNs. Specifically, the hybrid model combining frozen Provg-Gigapath features with ResNet50 pretrained on ImageNet achieved a balanced accuracy of 0.83, representing a reduction in performance relative to the best-performing fine-tuned CNN architectures. These findings collectively suggest that task-specific fine-tuning of CNNs for cellular-level downstream applications yields superior results when compared to both hybrid and foundation model approaches.\hfill\break

Detailed results on more ablation studies and other dataset benchmarks including AtNorm-BR and AtNorm-MD \cite{ami-br} datasets will be included in the extended version of this article.

\section*{Discussion}

Our proposed approach leverages a ResNet-based backbone with specialized heads to address class imbalance and annotator consensus—enabling robust classification of mitotic figure subtypes and their difficulty. This architecture ensures efficient, high-performance discrimination even under challenging, imbalanced, and discordant annotation scenarios. 

The results demonstrate the robustness and generalizability of the proposed deep learning-based approach for differentiating atypical mitotic figures (AMFs) from normal mitotic figures (NMFs) across a diverse set of tumor types and imaging conditions. Our approach consistently achieved high balanced accuracy during both stratified 5-fold cross-validation and preliminary leaderboard evaluation underscoring its reliability for cellular-level classification tasks. The overall area under the receiver operating characteristic curve (ROC AUC) further highlights the model's strong discriminative capability.  

Beyond technical metrics, the clinical utility of automated classification between atypical and normal mitotic figures is substantial. Accurate identification of AMFs can provide valuable prognostic information, as the presence and proportion of atypical mitoses are correlated with tumor aggressiveness and potential patient outcomes. By reducing reliance on subjective manual annotation and enabling high-throughput, standardized evaluation of histopathological slides, automated systems such as the one described here have the potential to facilitate more precise risk stratification, inform treatment decisions, and improve prognostic assessments in routine clinical practice. Moreover, robust generalization across inter-domain variability reinforces the promise of such methods for deployment in heterogeneous real-world settings, ultimately supporting improved diagnostic consistency and patient care. 

\begin{acknowledgements}
The authors thank the organizers of MIDOG-2025 for organizing the challenge and sharing the data with the community.
\end{acknowledgements} \hfill\break

\section*{Bibliography}
\bibliography{literature}

\begin{thebibliography}{11}
\providecommand{\natexlab}[1]{#1}
\providecommand{\url}[1]{\texttt{#1}}
\expandafter\ifx\csname urlstyle\endcsname\relax
  \providecommand{\doi}[1]{doi: #1}\else
  \providecommand{\doi}{doi: \begingroup \urlstyle{rm}\Url}\fi

\bibitem[Batistatou(2004)]{mitosis}
A~Batistatou.
\newblock Mitoses and cancer.
\newblock \emph{Med Hypotheses}, 63:\penalty0 281--2, 2004.

\bibitem[Lashen et~al.(2022)Lashen, Toss, Alsaleem, Green, Mongan, and
  Rakha]{AtypMitBC}
A~Lashen, M.S Toss, M~Alsaleem, A.R Green, N.P Mongan, and E~Rakha.
\newblock The characteristics and clinical significance of atypical mitosis in
  breast cancer.
\newblock \emph{Mod Pathol}, 35:\penalty0 1341--1348, 2022.

\bibitem[Dawson(2023)]{DigPathMitosis}
H~Dawson.
\newblock Digital pathology - rising to the challenge.
\newblock \emph{Front Med (Lausanne)}, 10:\penalty0 1180693, 2023.

\bibitem[Jiménez and Racoceanu(2019)]{Limitations}
G~Jiménez and D~Racoceanu.
\newblock Deep learning for semantic segmentation vs. classification in
  computational pathology: Application to mitosis analysis in breast cancer
  grading.
\newblock \emph{Front Bioeng Biotechnol}, 7:\penalty0 145, 2019.

\bibitem[Veta et~al.(2019)Veta, Heng, Stathonikos, Bejnordi, Beca, Wollmann,
  Rohr, Shah, Wang, Rousson, Hedlund, Tellez, Ciompi, Zerhouni, Lanyi, Viana,
  Kovalev, Liauchuk, Phoulady, Qaiser, Graham, Rajpoot, Sjöblom, Molin, Paeng,
  Hwang, Park, Jia, Chang, Xu, Beck, van Diest, and Pluim]{TUPAC16}
M~Veta, Y.J Heng, N~Stathonikos, B.E Bejnordi, F~Beca, T~Wollmann, K~Rohr, M.A
  Shah, D~Wang, M~Rousson, M~Hedlund, D~Tellez, F~Ciompi, E~Zerhouni, D~Lanyi,
  M~Viana, V~Kovalev, V~Liauchuk, H.A Phoulady, T~Qaiser, S~Graham, N~Rajpoot,
  E~Sjöblom, J~Molin, K~Paeng, S~Hwang, S~Park, Z~Jia, E.I Chang, Y~Xu, A.H
  Beck, P.J van Diest, and J.P.W Pluim.
\newblock Predicting breast tumor proliferation from whole-slide images: The
  tupac16 challenge.
\newblock \emph{Med Image Anal}, 54:\penalty0 111--21, 2019.

\bibitem[mit(2014)]{mitos}
Mitos-atypia-14 - grand challenge.
\newblock \url{https://mitos-atypia-14.grand-challenge.org/}, 2014.

\bibitem[M et~al.(2023)M, Stathonikos, Bertram, Klopfleisch, Ter~Hoeve, Ciompi,
  Wilm, Marzahl, Donovan, Maier, Breen, Ravikumar, Chung, Park, Nateghi,
  Pourakpour, Fick, Ben~Hadj, Jahanifar, Shephard, Dexl, Wittenberg, Kondo,
  Lafarge, Koelzer, Liang, Wang, Long, Liu, Razavi, Khademi, Yang, Wang, Erber,
  Klang, Lipnik, Bolfa, Dark, Wasinger, Veta, and Breininger]{midog2021}
Aubreville. M, N~Stathonikos, C.A Bertram, R~Klopfleisch, N~Ter~Hoeve,
  F~Ciompi, F~Wilm, C~Marzahl, T.A Donovan, A~Maier, J~Breen, N~Ravikumar,
  Y~Chung, J~Park, R~Nateghi, F~Pourakpour, R.H.J Fick, S~Ben~Hadj,
  M~Jahanifar, A~Shephard, J~Dexl, T~Wittenberg, S~Kondo, M.W Lafarge, V.H
  Koelzer, J~Liang, Y~Wang, X~Long, J~Liu, S~Razavi, A~Khademi, S~Yang, X~Wang,
  R~Erber, A~Klang, K~Lipnik, P~Bolfa, M.J Dark, G~Wasinger, M~Veta, and
  K~Breininger.
\newblock Mitosis domain generalization in histopathology images - the midog
  challenge.
\newblock \emph{Med Image Anal}, 84:\penalty0 102699, 2023.

\bibitem[Aubreville et~al.(2022)Aubreville, Bertram, Breininger, Jabari,
  Stathonikos, and Veta]{midog2022}
M~Aubreville, C~Bertram, K~Breininger, S~Jabari, N~Stathonikos, and M~Veta.
\newblock Mitosis domain generalization challenge 2022.
\newblock In \emph{25th International Conference on Medical Image Computing and
  Computer Assisted Intervention (MICCAI 2022)}, 2022.
\newblock \doi{10.5281/zenodo.6362337}.

\bibitem[Ammeling et~al.(2025)Ammeling, Aubreville, Banerjee, Bertram,
  Breininger, Hirling, Horvath, Stathonikos, and Veta]{MIDOG2025}
J~Ammeling, M~Aubreville, S~Banerjee, C.A Bertram, K~Breininger, D~Hirling,
  P~Horvath, N~Stathonikos, and M~Veta.
\newblock Mitosis domain generalization challenge 2025.
\newblock \url{https://doi.org/10.5281/zenodo.15077361}, 2025.

\bibitem[Aubreville et~al.(2023)Aubreville, Wilm, Stathonikos, Breininger,
  Donovan, Jabari, Veta, Ganz, Ammeling, van Diest, et~al.]{midog++}
M~Aubreville, F~Wilm, N~Stathonikos, K~Breininger, T.A Donovan, S~Jabari,
  M~Veta, J~Ganz, J~Ammeling, P.J van Diest, et~al.
\newblock A comprehensive multi-domain dataset for mitotic figure detection.
\newblock \emph{Scientific data}, 10\penalty0 (1):\penalty0 484, 2023.

\bibitem[Bertram et~al.(2025)Bertram, Weiss, Donovan, Banerjee, Conrad,
  Ammeling, Klopfleisch, Kaltenecker, and Aubreville]{ami-br}
Christof~A Bertram, Viktoria Weiss, Taryn~A Donovan, Sweta Banerjee, Thomas
  Conrad, Jonas Ammeling, Robert Klopfleisch, Christopher Kaltenecker, and Marc
  Aubreville.
\newblock Histologic dataset of normal and atypical mitotic figures on human
  breast cancer (ami-br).
\newblock In \emph{BVM Workshop}, pages 113--118. Springer, 2025.

\end{thebibliography}

\end{document}